\documentclass[aps,prl,twocolumn,showpacs,superscriptaddress,groupedaddress,nofootinbib]{revtex4}
\usepackage{amsmath,amsfonts,amssymb,array}
\usepackage{bm,bbm,bbm,bbold}
\usepackage{enumerate}
\usepackage{dsfont}
\usepackage{float}
\usepackage{graphicx}
\usepackage{longtable}
\usepackage{multirow}
\usepackage{slashed,subfigure}
\usepackage{xcolor}
\usepackage[colorlinks = true,linkcolor = blue]{hyperref} 
\definecolor{darkgreen}{rgb}{0.0, 0.5, 0.0}
\hypersetup{bookmarks=true,unicode=true,pdftoolbar=true,pdfmenubar=true,
  pdffitwindow=false,pdfstartview={FitH},
  pdftitle={heavyNThreshold},pdfauthor={Ruiz,Spannowsky,Waite},
  pdfsubject={Subject},pdfcreator={Creator},pdfproducer={Producer},
  pdfkeywords={Heavy Neutrinos}{Gluon Fusion}{Resummation},
  pdfnewwindow=true,colorlinks=true,
  linkcolor=blue,citecolor=purple,filecolor=magenta,urlcolor=cyan}
\providecommand{\tabularnewline}{\\}

\newcommand{\invfb}{{\rm ~fb^{-1}}}

\def\GeV{{\rm ~GeV}}
\def\TeV{{\rm ~TeV}}
\def\eff{{\rm eff}}

\newcommand{\nkll}[1]{{\rm N}^{#1}{\rm LL}}
\newcommand{\nklo}[1]{{\rm N}^{#1}{\rm LO}}

\newcommand{\confirm}{\textcolor{black}}
\begin{document}
\leftline{}
\rightline{IPPP/17/48}

\title{Heavy neutrinos from gluon fusion}

\author{Richard Ruiz}  
\email{richard.ruiz@durham.ac.uk}
\affiliation{Institute for Particle Physics Phenomenology, Department of Physics, Durham University, Durham DH1 3LE, U.K.}

\author{Michael Spannowsky}  
\email{michael.spannowsky@durham.ac.uk}
\affiliation{Institute for Particle Physics Phenomenology, Department of Physics, Durham University, Durham DH1 3LE, U.K.}

\author{Philip Waite}  
\email{p.a.waite@durham.ac.uk}
\affiliation{Institute for Particle Physics Phenomenology, Department of Physics, Durham University, Durham DH1 3LE, U.K.}

\date{\today}

\begin{abstract}
Heavy neutrinos, a key prediction of many standard model extensions, remain some of the most searched-for objects at collider experiments.
In this context, we revisit the premise that the gluon fusion production mechanism, $gg \to Z^*/h^* \to N\nu_\ell$,
is phenomenologically irrelevant at the CERN LHC and report the impact of soft gluon corrections to the production cross section.
We resum threshold logarithms up to next-to-next-to-next-to-leading logarithmic accuracy (N$^3$LL),
thus capturing the dominant contributions to the inclusive cross section up to next-to-next-to-leading order (N$^2$LO).
For $m_N > 150$ GeV and collider energies $\sqrt{s} = 7 - 100$ TeV, corrections to the Born rates span $+160$ to $+260\%$.
At $\sqrt{s}$=14 TeV, the resummed channel is roughly equal in size to the widely-believed-to-be-dominant charged current Drell-Yan process 
and overtakes it outright at $\sqrt{s} \gtrsim 20-25$ TeV. 
Results are independent of the precise nature/mixing of $N$ and hold generically for other low-scale seesaws.
Findings are also expected to hold for other exotic leptons and broken axial-vector currents,
particularly as the $Z^*$ contribution identically reduces to that of a pseudoscalar.
\end{abstract}

\maketitle

\section{Introduction}\label{sec:Intro}
In stark contrast to the standard model (SM) of particle physics, 
neutrinos have  nonzero mass, and  misaligned flavor and mass eigenstates~\cite{Ahmad:2002jz, Ashie:2005ik}. 
Hence, the origins of their sub-eV masses and large mixing angles are two of the most pressing open questions in particle physics today. 
In light of recent evidence for the Higgs mechanism's role in generating charged lepton masses~\cite{Aad:2015vsa,Chatrchyan:2014nva}, 
we argue that the existence of neutrino Dirac masses comparable to other elementary fermions' Dirac masses is an increasingly likely prospect. 
If this is the case, then observed neutrino phenomenology can be accommodated by low-scale seesaw mechanisms, 
such as the Inverse~\cite{Mohapatra:1986aw,Mohapatra:1986bd,Bernabeu:1987gr} or Linear~\cite{Akhmedov:1995ip,Akhmedov:1995vm} seesaw models.

In such seesaw scenarios, TeV-scale heavy neutrinos' mass eigenstates ($N$) can couple to electroweak (EW) bosons with sizable ~\cite{Antusch:2014woa,Fernandez-Martinez:2016lgt} active-sterile mixing, but at the same time do not decouple from Large Hadron Collider (LHC) phenomenology~\cite{Pilaftsis:1991ug,Kersten:2007vk} due to their pseudo-Dirac nature. Subsequently, low-scale seesaw mechanisms can be tested at the LHC with $\mathcal{O}(100-1000)\invfb$~\cite{Arkani-Hamed:2015vfh,Golling:2016gvc,Antusch:2016ejd,Arganda:2015ija,Arganda:2014dta,Baglio:2016bop}, demonstrating the sensitivity and complementarity of collider and oscillation experiments.

\begin{figure*}[!th]
\includegraphics[width=0.81\textwidth]{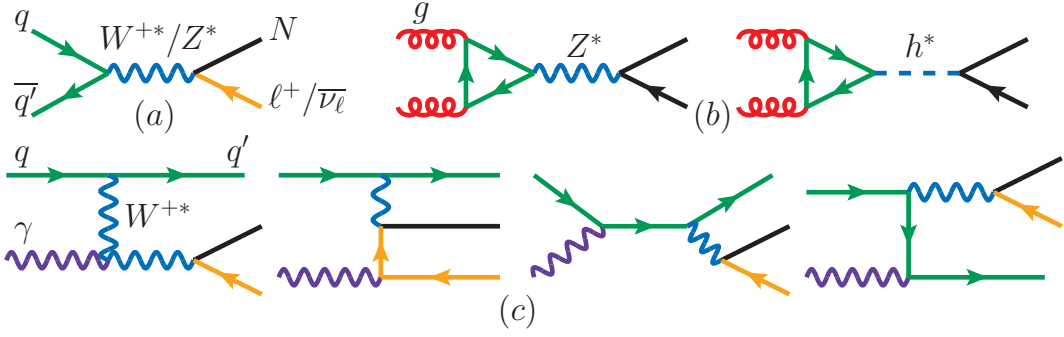}	
\caption{
Born diagrams for heavy $N$ production via the (a) DY, (b) GF, and (c) VBF processes. 
Drawn using JaxoDraw~\cite{Binosi:2003yf}.
}
\label{fig:feynman}
\end{figure*}

Hadron collider investigations of heavy $N$ typically rely on the charged current (CC) Drell-Yan (DY) process~\cite{Keung:1983uu},  
shown in Fig.~\ref{fig:feynman}(a) and given by
\begin{equation}
 q \overline{q'} \to W^{\pm(*)} \to N \ell^\pm, \quad q\in\{u,c,d,s,b\},
 \label{eq:ccdy}
\end{equation}
or the sizable vector boson fusion (VBF) channel~\cite{Datta:1993nm,Dev:2013wba,Alva:2014gxa,Degrande:2016aje},
\begin{equation}
 q \gamma \overset{W \gamma \to N \ell}{\longrightarrow} N \ell^\pm ~q'.
 \label{eq:vbf}
\end{equation}
As seen in Fig.~\ref{fig:feynman}(c), VBF is driven by the $W\gamma\to N\ell$ subprocess
and receives longitudinal $W$ enhancements for $m_N\gg M_W$~\cite{Alva:2014gxa}.
Notable is the renewed interest~\cite{Hessler:2014ssa,Degrande:2016aje} 
in the gluon fusion (GF) process~\cite{Degrande:2016aje,Willenbrock:1985tj,Dicus:1991wj,Hessler:2014ssa}, shown in Fig.~\ref{fig:feynman}(b),
\begin{equation}
  g g \to Z^* / h^* \to N  \overset{(-)}{\nu_\ell}.
  \label{eq:gf}
\end{equation}
{Variants of this process have been 
studied recently in}~\cite{BhupalDev:2012zg,Batell:2015aha,Gago:2015vma,Accomando:2016rpc,Nemevsek:2016enw,Caputo:2017pit,Das:2017zjc}.
While GF proceeds anomalously through off-shell $Z^*/h^*$ bosons
and is formally an $\mathcal{O}(\alpha_s^2)$ correction to the neutral current (NC) DY process,
$q \overline{q} \to Z^* \to N \overset{(-)}{\nu_\ell},$
the channel's cross section is known to surpass the DY and VBF rates for collider energies 
$\sqrt{s} \gtrsim 30-40$ TeV~\cite{Willenbrock:1985tj,Hessler:2014ssa,Degrande:2016aje}.
At 14 TeV, GF is factors smaller than the DY channels.
These conclusions are noteworthy, as they rely on the GF rate at
leading-order (LO) accuracy being a good estimate of the total  cross section.
However, at LO, the GF rates for the SM Higgs boson~\cite{Dawson:1990zj,Spira:1995rr,Spira:1997dg,Harlander:2002wh,Ravindran:2003um,Catani:2014uta,Anastasiou:2015ema}, 
heavy scalars~\cite{Spira:1997dg}, and pseudoscalars~\cite{Harlander:2002vv,Anastasiou:2002wq,Ahmed:2016otz} are greatly underestimated.

In light of this, we report, for the first time, the impact of soft gluon corrections to heavy $N$ production in GF.
We resum threshold logarithms up to next-to-next-to-next-to-leading logarithmic accuracy (N$^3$LL).
For GF, this captures the leading contributions to the inclusive cross section $(\sigma)$ up to next-to-next-to-leading order (N$^2$LO)~\cite{Bonvini:2014qga}. 
Our findings have immediate impact on searches at hadron colliders,
and thereby challenge the paradigm that GF is phenomenologically irrelevant for the discovery and study of heavy $N$ at the LHC.

For $m_N = 150-1000\GeV$, $\sqrt{s} = 7 - 100$ TeV, and scale choices comparable to the hard process, we report
\begin{eqnarray}
K^{\nkll{3}} = \sigma^{\nkll{3}}/\sigma^{\rm LO} &:&  \confirm{2.6-3.6},
  \\
K^{\nkll{2}} = \sigma^{\nkll{2}}/\sigma^{\rm LO} &:&  \confirm{2.3-3.0}.
\end{eqnarray}
We find that GF dominates over the DY-like processes of Eq.~(\ref{eq:ccdy}) 
for \confirm{$m_N = 500-1000$} GeV at $\confirm{\sqrt{s} \gtrsim 20-25}$ TeV.
The corrections exhibit perturbative convergence and are consistent with those for Higgs and heavy (pseudo)scalar 
production~\cite{Dawson:1990zj,Spira:1995rr,Harlander:2002wh,Ravindran:2003um,Catani:2014uta,Spira:1997dg,Harlander:2002vv,Anastasiou:2002wq,Anastasiou:2015ema,Ahmed:2016otz}.
Our results are independent of the precise nature/mixing of $N$ and hold generically for other low-scale seesaws.
Results are also expected to hold for other exotic leptons, e.g. triplet leptons in the Type III seesaw and other colorless, axial-vector currents.

This report continues as follows:
We first describe our phenomenological heavy $N$ model,
then present the resummation formalism employed, emphasizing a new treatment of the $Z^*$ current.
After summarizing our computational setup, we present our results and conclude.

\section{Heavy Neutrino Model}\label{sec:theory}
Throughout this study, we adopt the neutrino mixing formalism of~\cite{Atre:2009rg}:
For $i~(m) =1,\dots,3$ left-handed (light) states and $j~(m')=1,\dots,n$ right-handed (heavy) states,
chiral neutrinos can be rotated into mass eigenstates by
\begin{eqnarray}
\begin{pmatrix} \nu_{Li} \\  N_{Rj}^c \end{pmatrix}
=
\begin{pmatrix} 
U_{3\times3} && V_{3\times n} \\
X_{n\times3} && Y_{n\times n}
\end{pmatrix}
\begin{pmatrix} \nu_{m} \\ N_{m'}^c \end{pmatrix}.
\end{eqnarray}
After further rotating the charged leptons into the mass basis, the flavor state $\nu_{\ell}$ in the mass basis is explicitly
\begin{equation}
 \nu_{\ell} = \sum_{m=1}^{3} U_{\ell m}\nu_{m} + \sum_{m'=1}^{n}V_{\ell m'} N^{c}_{m'}.
\end{equation}
$U_{\ell m}$ is the observed light neutrino mixing matrix and $V_{\ell m'}$ parametrizes
active-heavy mixing. For EW-scale $N_{m'}$, the latter is constrained by precision EW data to be $\vert V_{\ell N}\vert \lesssim 10^{-2}-10^{-1}$~\cite{Antusch:2014woa,Fernandez-Martinez:2016lgt}.
For simplicity, we consider only the lightest heavy state, denoted $N$.

In the mass basis, the EW interaction Lagrangian is
{
\small{
\begin{eqnarray}
  \mathcal{L}_{\rm Int.} = 
  &-& \frac{g}{\sqrt{2}}W^+_\mu		\sum_{\ell=e}^\tau \sum_{m=1}^3 ~\overline{\nu_m} ~U_{\ell m}^*	~\gamma^\mu P_L\ell^-\nonumber\\
  &-& \frac{g}{\sqrt{2}}W^+_\mu		\sum_{\ell=e}^\tau 		~\overline{N^c} ~V_{\ell N}^* 	~\gamma^\mu P_L\ell^-\nonumber\\
    &-& \frac{g}{2\cos\theta_W}Z_\mu	\sum_{\ell=e}^\tau \sum_{m=1}^3	~\overline{\nu_m} ~U_{\ell m}^*	~\gamma^\mu P_L\nu_\ell\nonumber
  \end{eqnarray}
\begin{eqnarray}
  &-& \frac{g}{2\cos\theta_W}Z_\mu	\sum_{\ell=e}^\tau 		~\overline{N^c} ~V_{\ell N}^*	~\gamma^\mu P_L\nu_\ell\nonumber\\
  &-& \frac{g m_N}{2M_W} h	\sum_{\ell=e}^\tau 		~\overline{N^c} ~V_{\ell N}^*	 P_L\nu_\ell + \text{H.c.}
  \label{eq:Lagrangian}
\end{eqnarray}
}
}
At the production level, $\vert V_{\ell N}\vert$ factorizes out of cross sections, a result that holds at all orders in $\alpha_s$~\cite{Ruiz:2015zca,Degrande:2016aje}. 
This allows one to define~\cite{Han:2006ip} a ``bare'' cross section $\sigma_{0}$ in which one sets $\vert V_{\ell N}\vert=1$.
Subsequently, flavor model-independent cross sections are given by 
\begin{equation}
 \sigma(pp\rightarrow N+X) /  \vert V_{\ell N}\vert^2 ~=~ \sigma_{0}(pp\rightarrow N+X).
\end{equation}
Due to this factorization, the QCD corrections we present are universal across low-scale seesaws that feature $N$.

\section{Threshold Resummation Formalism}\label{sec:scet}
We now summarize our resummation formalism and the special consideration of the $Z^*$ mediator in GF.

For a color-singlet final state $V$,
the inclusive $pp\to V+X$ fixed-order (FO) cross section 
is given generically by the Collinear Factorization Theorem\footnote{The equivalent measure: 
$\int_{\tau_0}^1 d\xi_1 \int_{\tau_0/\xi_1}^1 dx  \int_{x}^1 dz/z$, with $\xi_2 = x/z$,
may lead to faster numerical convergence for some processes.}~\cite{Collins:1984kg,Collins:1985ue,Collins:2011zzd}, 
\small{
\begin{eqnarray}
\label{eq:resummed_xsec}
\sigma^{\rm FO} &=& f\otimes f\otimes \Delta\otimes \hat{\sigma}
\\
&=& 
\frac{1}{1+\delta_{ij}}\sum_{i,j,\beta} \int_{\tau_0}^1d\tau \int_\tau^1 \frac{d\xi_1}{\xi_1} \int_{\tau/\xi_1}^1\frac{dz}{z} 
\nonumber\\
& &
\left[f_{i/p}(\xi_1)f_{j/p}(\xi_2) + (1\leftrightarrow2)\right] 
\Delta^{\beta,{\rm FO}}_{ij}(z) ~\hat{\sigma}_{ij}^\beta. \qquad
\label{eq:cft}
\end{eqnarray}
}\normalsize
That is, the hadronic scattering rate $\sigma$ is the 
convolution $(\otimes)$ of parton distribution functions (PDFs) $f$, the soft coefficient function $\Delta$,
and the partonic-level $ij\to V$ hard scattering rate $\hat{\sigma}$, which occurs at the hard scattering scale $Q = \sqrt{p_V^2}$.
Scale dependence of these quantities is implied but made explicit below.
$f_{i/p}(\xi,\mu_f)$ are the likelihoods of observing parton $i$ in $p$ carrying longitudinal momentum $p_z^i = \xi p_z^p\gg p_T^i$, 
when DGLAP-evolved~\cite{Gribov:1972ri, Dokshitzer:1977sg,Altarelli:1977zs} to a factorization scale $\mu_f$,
generating the partonic scale $\sqrt{\hat{s}} = \sqrt{\xi_1 \xi_2 s}$.
$\Delta^{\beta,{\rm FO}}_{ij}(z) = \delta(1-z) + \mathcal{O}(\alpha_s)$ accounts for soft gluons carrying a 
momentum fraction $(1-z)$, with $z=Q^2/\hat{s}$,
emitted in the $ij\to A$ transition $(\hat{\sigma}^\beta)$ via a color/Lorentz structure labeled as $\beta$.
Above, $\tau_0 = \min\{Q^2\}/s$ is the kinematic threshold below which $ij\to A$ is kinematically forbidden,
and $\tau = Q^2/s = \xi_1 \xi_2 z$ is similarly the hard threshold.

For the $gg\to Z^*/h^* \to N\nu_\ell$ process, with $\beta\in\{Z,h\}$, $Q^2=(p_N + p_\nu)^2 > m_h^2,M_Z^2$ and $\tau_0 \approx m_N^2 / s$,
the hard partonic-level Born cross sections are\footnote{We note that the expression for $\hat{\sigma}^h$ in~\cite{Hessler:2014ssa} contains typographic errors.}~\cite{Willenbrock:1985tj,Hessler:2014ssa}
\begin{eqnarray}
 \hat{\sigma}^Z &=& G_F^2 \frac{\alpha_s^2(\mu_r)\vert V_{\ell N}\vert^2  }{2^4(4\pi)^3} m_N^2(1-r_N)^2 \vert F_Z(Q^2) \vert^2,
 \label{eq:sigmaZ}
 \\
 \hat{\sigma}^h &=& G_F^2 \frac{\alpha_s^2(\mu_r)\vert V_{\ell N}\vert^2  }{2^4(4\pi)^3} 
 \frac{m_N^2 Q^4 (1-r_N)^2}{(Q^2-m_h^2)^2} \vert F_h(Q^2) \vert^2,\qquad 
 \label{eq:sigmaH}
\end{eqnarray}
where $r_X = m_X^2 / Q^2$. 
For quarks with weak isospin charge $(T_L^3)_q = \pm1/2$, the $Z/h$ one-loop form factors are
\begin{eqnarray}
 F_Z(Q^2) &=& \sum_{q=t,b\dots} 2(T_L^3)_q ~ \left[1-2 r_q f(r_q)\right], 
 \\
 F_h(Q^2) &=& \sum_{q=t,b\dots} 2 r_q \left[2  + (1 - 4r_q)f(r_q)\right], ~\text{with},
 \\
  f(r) &=& 
\left\{\begin{matrix}
2\left(\sin^{-1} \frac{1}{2\sqrt{r}}\right)^2,	 & r > \frac{1}{4}, \\ 
-\frac{1}{2}\left[\log\left(\frac{1+\sqrt{1-4r}}{1-\sqrt{1-4r}}\right)-i\pi\right]^2,  & r \leq \frac{1}{4}.
\end{matrix}\right.\quad
\label{eq:loopFn}
\end{eqnarray}

A few remarks:
$(i)$ While we use the fully integrated $\hat{\sigma}^{\beta}$, the resummation formalism we employ~\cite{Becher:2006mr,Becher:2006nr} operates in momentum space. Hence, phase-space cuts on an $n$-body final state can be implemented if one starts from the differential $d\hat{\sigma}^{\beta}$.
$(ii)$ The $Z^*/h^*$ contributions add incoherently due to the \mbox{(anti)}symmetric nature of the $(Z^*)h^*$ coupling~\cite{Willenbrock:1985tj}.
$(iii)$ The similarity of $\hat{\sigma}^{Z}$ and $\hat{\sigma}^{h}$ follows from the fact that, after summing over SU$(2)_L$ doublet constituents, the net $Z^*$ contribution is a pseudoscalar-like coupling proportional to quark Yukawa couplings. This is in accordance with the Goldstone Equivalence Theorem, pointed out first for the $gg\to N\nu_\ell$ process in Ref.~\cite{Degrande:2016aje}. Subsequently, this asymptotic behavior means one can further simplify the original expressions of Refs.~\cite{Willenbrock:1985tj,Hessler:2014ssa} to those above.

The axial-vector-pseudoscalar correspondence, however, is more general:
For a massive, colorless vector $V(q_\mu)$ participating in the loop process $gg\to V^{*}$,
the most general current-propagator contraction (in the unitary gauge) is of the form 
$\Gamma^\mu \Pi_{\mu\nu}\sim (g_V\gamma^\mu + g_A\gamma^\mu\gamma^5)(g_{\mu\nu} - q_\mu q_\nu / M_V^2)$.
By $C$-symmetry (Furry's theorem), the vector current $g_V\gamma^\mu$ vanishes;
by angular momentum conservation (Landau-Yang theorem), the transverse polarization $g_{\mu\nu}$ does not contribute.
Hence, $\Gamma^\mu \Pi_{\mu\nu}\sim g_A\gamma^\mu\gamma^5 q_\mu q_\nu / M_V^2$.
After decomposing quark propagators in the triangle loop via spinor completeness relations
and exploiting the Dirac equation, one finds  $\Gamma^\mu \Pi_{\mu\nu}\sim \gamma^5(2m_f q_\nu / M_V^2)$.
That is, a pseudoscalar coupling proportional to the quark mass $m_f$.

Moreover, emissions of soft gluons {off fermions} do not change the loop's structure due to soft factorization. Therefore, one may approximate soft QCD corrections to the $gg\to V^*$ subprocess for  $V^*$ possessing axial-vector couplings to fermions with those corrections for a pseudoscalar. This is a main finding of this work and was not observed in previous resummations of $gg\to Z^*$.

As $Q$ approaches the partonic threshold $\sqrt{\hat{s}}$, 
accompanying gluon radiation is forced to be soft, with $E_g \sim \sqrt{\hat{s}}(1-z)$.
This generates numerically large phase-space logarithms
of the form $\log(1-z)$ that spoil the perturbative convergence of Eq.~(\ref{eq:cft}).
Threshold logarithms, however, factorize and can be resummed 
to all orders in ${\alpha_s\log(1-z)}$ via exponentiation~\cite{Catani:1989ne,Sterman:1986aj,Catani:1990rp,Contopanagos:1996nh,Forte:2002ni}.

We perform this resummation by working in the Soft-Collinear Effective Theory (SCET) framework~\cite{Bauer:2000yr,Bauer:2001yt,Beneke:2002ph}.
This permits Eq.~(\ref{eq:cft}) to be factorized directly in momentum space~\cite{Becher:2006mr,Becher:2006nr}
by segmenting and regulating divergent regions of phase space with hard and soft scales, $\mu_h$ and $\mu_s$.
(This is unlike perturbative QCD where one works in Mellin space~\cite{Catani:1989ne,Sterman:1986aj,Catani:1990rp}.)
Scale invariance of physical observables then implies that factored components
can also be independently renormalization group (RG)-evolved and matched
via exponentiation~\cite{Catani:1989ne,Sterman:1986aj,Catani:1990rp,Contopanagos:1996nh,Forte:2002ni}.
Thus, numerically large quantities are replaced with perturbative ones regulated by $\mu_{h}$ and $\mu_{s}$
and with RG-evolution coefficients that run $\mu_{h}$ and $\mu_{s}$ to  $\mu_f$ and $Q$.

In practice, the resummation procedure reduces to replacing the soft coefficient function $\Delta^{\beta,{\rm FO}}_{ij}(z)$ in Eq.~(\ref{eq:cft}):
\begin{equation}
 \sigma^{\rm FO}\to\sigma^{\rm Res} \quad:\quad \Delta^{\beta,{\rm FO}}_{ij}(z) \to \Delta^{\beta,{\rm Res}}_{ij}(z).
\end{equation}
For GF production of heavy leptons via $s$-channel pseudoscalar $(\beta=Z)$ and scalar $(\beta=h)$ mediators, 
as given in Fig.~\ref{fig:feynman}(b), the SCET-based soft coefficient $\Delta_{gg}^{\beta,\text{Res}}$ 
in the notation of ~\cite{Becher:2007ty,Ahrens:2008nc} is
\small{
\begin{eqnarray}
\label{eq:res_coeff}
& & 
\Delta_{gg}^{\beta, \text{Res}}(z) = \lvert C_\beta(Q^2,\mu_h^2) \rvert^2 U(Q^2,\mu_\alpha^2,\mu_h^2,\mu_s^2,\mu_f^2) 
\quad\\
& & 
\times
\frac{\sqrt{z}~z^{-\eta}}{(1-z)^{1-2\eta}} 
\tilde{s}_{\mathrm{Higgs}}\left(\log\frac{Q^2(1-z)^2}{\mu_s^2z}+\partial_\eta,\mu_s\right) \frac{e^{-2\gamma_E\eta}}{\Gamma(2\eta)}.\nonumber
\label{eq:softCoeff}
\end{eqnarray}
}\normalsize
$C_\beta$ is the process-dependent, so-called hard function and accounts for (hard) virtual corrections to the hard process.
For $\beta=h$, the function is given by the two-step SCET matching coefficients $C_t$ and $C_{S}$ of ~\cite{Ahrens:2008nc}, with
\begin{eqnarray}
C_h(Q^2,\mu_h^2) &\equiv&  C_t(m_{t}^2,\mu_t^2) C_S(-Q^2,\mu_h^2), ~\text{and}\\
C_X(Q^2,\mu^2) &=& \sum_{n=0}^{\infty}C_{X}^{(n)}(Q^2,\mu^2) ~\left(\frac{\alpha_s(\mu)}{4\pi}\right)^n,\qquad
\end{eqnarray}
where $X\in\{t,S\}$. The product of $C_t$ and $C_S$, which can be expanded individually as power series in $\left(\alpha_s/4\pi\right)$,
is equivalent to a one-step SCET matching procedure when setting $\mu_t = \mu_h$~\cite{Berger:2010xi}.
For $\beta=Z$, the one-step matching hard function can also be expanded as a power series.
In the notation of~\cite{Ahmed:2015qpa,Ahmed:2016otz}, this is
\begin{eqnarray}
  C_Z(Q^2,\mu_h^2) &\equiv&  C_{g}^{A,\eff}(Q^2,\mu_h^2)  \nonumber\\
    &=& \sum_{n=0}^{\infty}C_{g,n}^{A,\eff}(Q^2,\mu_h^2) ~\left(\frac{\alpha_s(\mu_h)}{4\pi}\right)^n.\qquad
\end{eqnarray}
We briefly note that the $\log(\mu_h^2/m_{t}^2)$ term that appears in $C_{g,2}^{A,\eff}$ of ~\cite{Ahmed:2015qpa}
should be replaced with $\log(Q^2/m_{t}^2)$ in order to preserve the scale independence of the total cross section, a physical observable~\cite{Ahmed:2016otz}.
With this modification, both $C_{h}$ and $C_{Z}$ satisfy the evolution equation,
\begin{eqnarray}
 & & \frac{d}{d\log\mu}C_\beta(Q^2,\mu^2) =
 \nonumber\\
 & & \left[\Gamma^{A}_{\rm cusp}\log\left(\frac{Q^2}{\mu^2}\right) + \gamma^S + \gamma^t\right]C_\beta(Q^2,\mu^2),
\end{eqnarray}
for anomalous dimensions $\Gamma^{A}_{\rm cusp},\gamma^{S,t},$ as given in ~\cite{Becher:2007ty,Ahrens:2008nc}.

The soft scalar function $\tilde{s}_{\mathrm{Higgs}}$ describes (soft) radiation 
off incoming gluons and hence is universal for scalars and pseudoscalars.
The derivatives in $\tilde{s}_{\mathrm{Higgs}}$ are regular partial derivatives that act to the right,
before $\eta \equiv  2a_{\Gamma}(\mu_s^2,\mu_f^2)$ is evaluated numerically. 

We include an additional factor of $\sqrt{z}$ in Eq.~(\ref{eq:softCoeff}) with respect to~\cite{Becher:2007ty,Ahrens:2008nc}.
As noted in~\cite{Becher:2007ty,Ahrens:2008nc,Ahrens:2011mw,Broggio:2015lya}, the inclusion of the factor
accounts precisely for power corrections that are manifest in the traditional QCD/Mellin-space resummation formalism.
Numerically, we find this increases our total normalization by $\confirm{\mathcal{O}(5-10)\%}$ 
at $\nkll{3}$ and our residual scale dependence by $\confirm{\mathcal{O}(5\%)}$.

RG-running is described by the evolution function,
\vspace{0.1cm}
{\small{
\begin{align}
& U(Q^2,\mu_\alpha^2,\mu_h^2,\mu_s^2,\mu_f^2)  = 
\frac{\alpha_s^2(\mu_s^2)}{\alpha_s^2(\mu_f^2)}\left[
\frac{\beta(\alpha_s(\mu_s^2))/\alpha_s^2(\mu_s^2)}{\beta(\alpha_s(\mu_\alpha^2))/\alpha_s^2(\mu_\alpha^2)}\right]^2\qquad
\nonumber\\
&
\left(\frac{Q^2}{\mu_h^2}\right)^{-2a_{\Gamma}(\mu_h^2,\mu_s^2)}
\left \vert e^{4\mathcal{S}(\mu_h^2,\mu_s^2)-2a_{\gamma^{S}}(\mu_h^2,\mu_s^2)+4a_{\gamma^{B}}(\mu_s^2,\mu_f^2)} \right\vert,
\hspace{-1cm}
\label{eq:rgCoeff}
\end{align}
}}

\noindent
where $\mu_\alpha = \mu_{t}~(\mu_h)$ for $\beta=h~(Z)$.

For definitions and explicit expressions of the quantities in Eqs.~(\ref{eq:softCoeff})--(\ref{eq:rgCoeff}) up to $\mathcal{O}(\alpha_s^2)$, see~\cite{Ahrens:2008nc,Becher:2007ty}.
Mappings between $\nkll{k}$ accuracy and required ingredients can be found in ~\cite{Becher:2007ty,Bonvini:2014tea}. 
At $\nkll{3}$, one needs at two loops:
$C_\beta$ for both pseudoscalar~\cite{Ahmed:2015qpa} and scalar~\cite{Harlander:2000mg,Ahrens:2008nc},
as well as $\tilde{s}_{\mathrm{Higgs}}$~\cite{Idilbi:2006dg,Ahrens:2008nc,Becher:2007ty}.
Note that while the results of ~\cite{Ahmed:2015qpa,Harlander:2000mg,Ahrens:2008nc} are derived in the heavy top limit, 
$\mathcal{O}(Q^2/m_t^2)$ corrections to inclusive (pseudo)scalar cross sections are known to be $\mathcal{O}(1-10\%)$~\cite{Spira:1997dg,Caola:2011wq},
even for $Q^2\gg m_t^2$, justifying their use in our calculation.

\begin{figure*}[!th]
\subfigure{\includegraphics[width=0.48\textwidth]{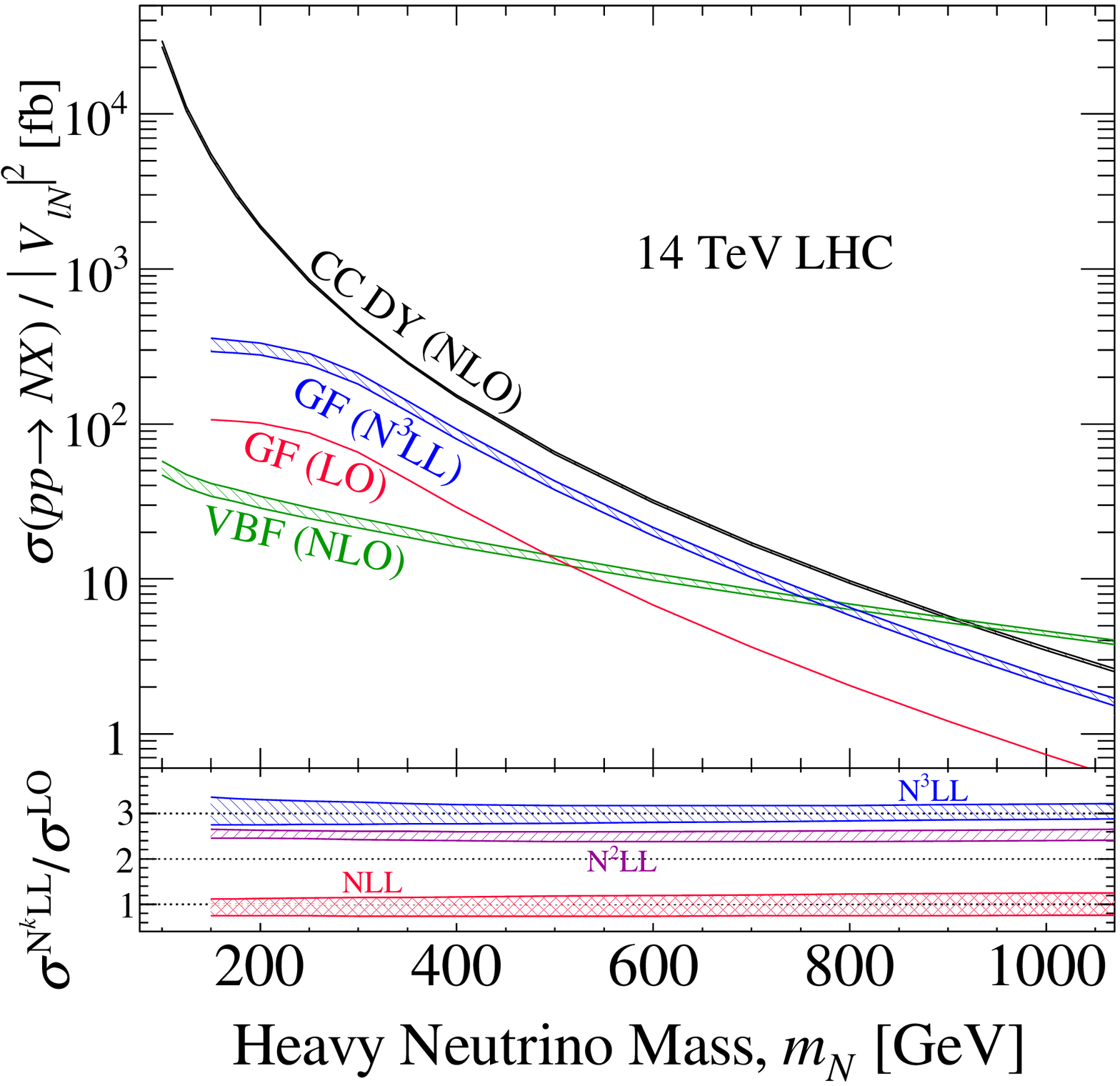}	\label{fig:xsecVsMass}	}
\subfigure{\includegraphics[width=0.48\textwidth]{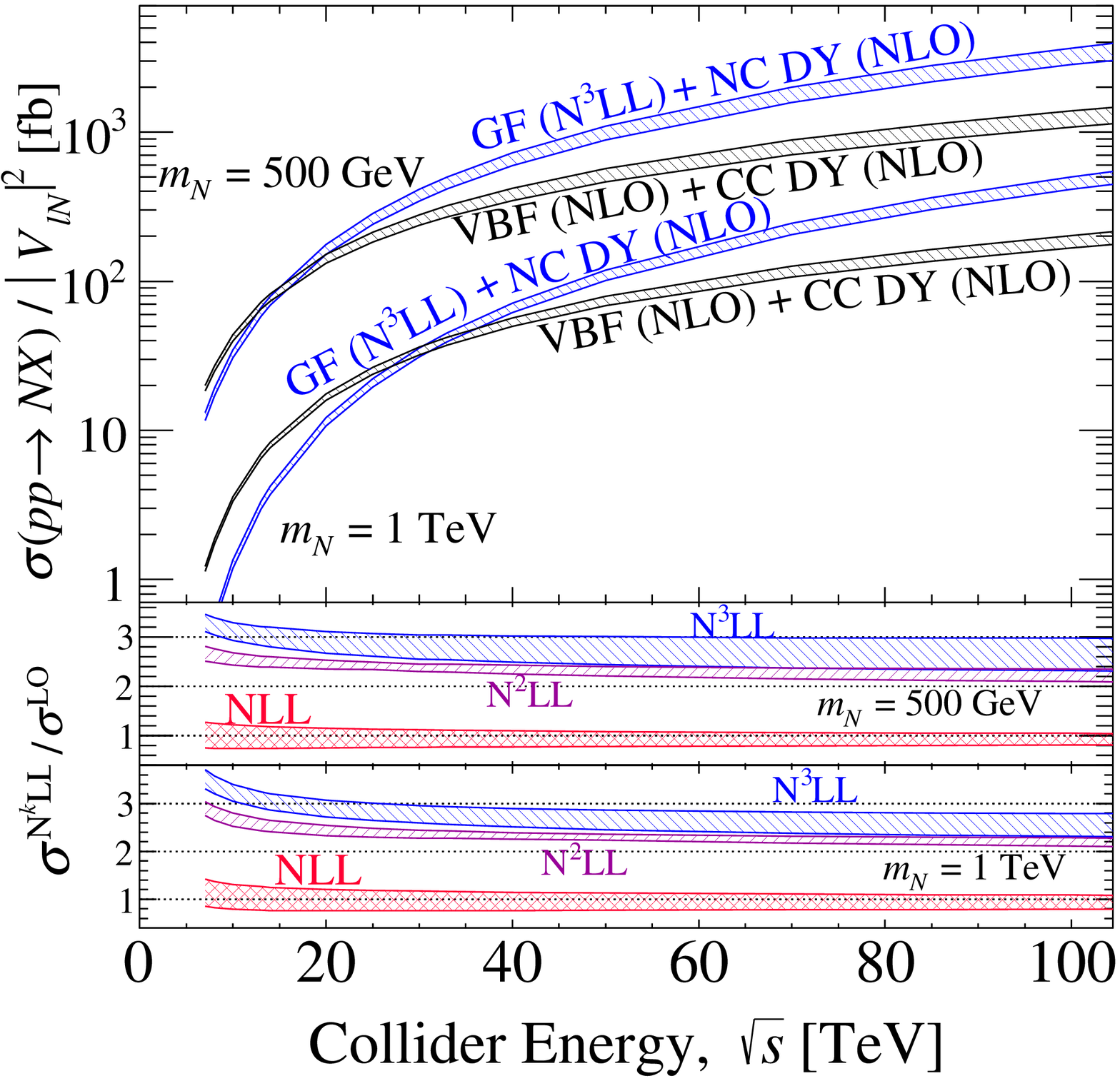}	\label{fig:xsecVsBeam}	}
\caption{Heavy $N$ cross sections [fb], divided by the mixing coefficient $\vert V_{\ell N}\vert^2$,
for various production mechanisms as a function of 
(L) neutrino mass $m_N$ [GeV] and (R)  collider energy [TeV]. 
Lower: Ratio of the resummed and Born GF predictions.
}
\label{fig:heavyNXSec}
\end{figure*}

\section{Computational Setup}\label{sec:mc}
For the DY and VBF channels,  we use the methodology of~\cite{Degrande:2016aje} to compute inclusive cross sections and uncertainties at NLO in QCD, but with the following exceptions:
we use the NNPDF 3.0 QED NLO PDF set~\cite{Ball:2014uwa,Ball:2013hta}
and do not apply phase-space cuts to the CC DY process. 
Scale choices and regulating VBF cuts are unchanged.
For GF, we adopt the additional SM inputs~\cite{Agashe:2014kda}:
\begin{eqnarray}
   m_b = 0\GeV, ~m_t = 173.2\GeV, ~m_h = 125.7\GeV. \nonumber
 \label{eq:smInputs}
\end{eqnarray}
To best match the accuracy of the resummation calculation, we use the NNPDF 3.0 NNLO+NNLL PDF set~\cite{Bonvini:2015ira};
while the set's uncertainties are sizable, the use of a FO PDF set would formally double-count initial-state gluons.
Cross sections are calculated using in-house code with Monte Carlo integration performed via the CUBA libraries~ \cite{Hahn:2004fe},
and checked at LO against~\cite{Hessler:2014ssa,Degrande:2016aje}.
The soft coefficient function $\Delta^{\beta,{\rm Res}}_{gg}$
is checked against~\cite{Bonvini:2014joa,Bonvini:2014tea,Ahmed:2016otz}.

To minimize the numerical impact of missing QCD corrections, we follow~\cite{Becher:2007ty,Ahrens:2008nc} and choose the scale scheme
\begin{equation}
 \mu_r, \mu_f, \mu_t, \mu_h = Q \quad\text{and}\quad \mu_s = Q(1-\tau)/(1+7\tau),
\end{equation}
{for both the Born and resummed GF calculations.}
For GF, we report the scale dependence associated with simultaneously varying 
$\mu_f,\mu_r,$ and $\mu_s$ over $0.5 \leq \mu_X/\mu_{\rm Default} \leq 2$.
While the $\mu_s$ dependence itself is numerically small,
we vary it jointly with $\mu_f$ to ensure that the subtraction terms required for numerical evaluation lead sufficiently to numerical convergence;
see~\cite{Ahrens:2008nc,Becher:2007ty} for more details.
Missing FO terms that would otherwise stabilize $\mu_f$ represents the largest source of uncertainty.
Indeed, we find other scale uncertainties to be relatively small owing to our high logarithmic accuracy.

In the following, we report only residual scale dependence.
For studies on PDF uncertainties in heavy $N$ production, see~\cite{Alva:2014gxa,Degrande:2016aje},
and for threshold-improved PDF uncertainties, see~\cite{Beenakker:2015rna,Mitra:2016kov}.
PDFs and $\alpha_s(\mu)$ are evaluated using the LHAPDF 6 libraries~\cite{Buckley:2014ana}.

\section{Results}\label{sec:results}
At $\sqrt{s} = 14\TeV$ and as a function of heavy $N$ mass, 
we show in Fig.~\ref{fig:heavyNXSec}(L) the inclusive $N$ production cross section (divided by active-heavy mixing $\vert V_{\ell N}\vert^2$)
for the CC DY and VBF processes at NLO, for GF at LO, andfor  GF at $\nkll{3}$.
The thickness of each curve corresponds to the residual scale dependence; no scale dependence is shown for GF at LO.
In the lower panel is the ratio of the resummed and Born GF rates.
We quantify the impact of QCD corrections with a $K$-factor, defined generically as
\begin{equation}
 K^{\nklo{j}+\nkll{k}} \equiv \sigma^{\nklo{j}+\nkll{k}} / \sigma^{\rm LO}.
\end{equation}

For $m_N = 150-1000$ GeV, cross sections span roughly
\begin{eqnarray}
\text{CC DY~NLO} 	&:&  \confirm{3.5~-5400~\mathrm{fb}}, \\
\text{GF~}\nkll{3}	&:&  \confirm{1.9~-280~\mathrm{fb}}, \\
\text{GF~LO} 		&:&  \confirm{0.73-110~\mathrm{fb}}, \\
\text{VBF~NLO}		&:&  \confirm{4.4~-37~\mathrm{fb}}. 
\end{eqnarray}
For GF, $K$-factors and uncertainties span approximately
\small{
\begin{eqnarray}
\text{GF~}\nkll{3}	&:&  K=\confirm{3.07 - 3.14} ~\text{with}~ \delta\sigma/\sigma=\confirm{\pm  8 - 13\%},\\
\text{GF~}\nkll{2}	&:&  K=\confirm{2.59 - 2.66} ~\text{with}~ \delta\sigma/\sigma=\confirm{\pm  6 -  9\%},\\
\text{GF~}{\rm NLL}	&:&  K=\confirm{1.00 - 1.06} ~\text{with}~ \delta\sigma/\sigma=\confirm{\pm 25 - 27\%}.\qquad
\end{eqnarray}
}\normalsize
These rates should be compared with DY (VBF) $K$-factors of $K^{\rm NLO}_{\rm DY~(VBF)} = 1.15-1.25~(0.98-1.06)$ 
and uncertainties of $(\delta\sigma/\sigma)^{\rm DY~(VBF)} = \pm 1-5~(5-11)\%$~\cite{Degrande:2016aje}.

{\small
\begin{table*}[!th]
\begin{tabular}{ c || c | c | c | c || c | c | c | c || c | c | c | c}
$\sqrt{s}$ & \multicolumn{4}{c||}{14 TeV} &\multicolumn{4}{c||}{33 TeV} &\multicolumn{4}{c}{100 TeV} \tabularnewline\hline
$m_N$ &\multicolumn{2}{c|}{300 GeV}&\multicolumn{2}{c||}{500 GeV}
      &\multicolumn{2}{c|}{300 GeV}&\multicolumn{2}{c||}{500 GeV}
      &\multicolumn{2}{c|}{300 GeV}&\multicolumn{2}{c}{1 TeV} \tabularnewline\hline
$\sigma ~/~ \vert V_{\ell N}\vert^2 ~$ [fb]	& $\sigma$ [fb]  & $K$ & $\sigma$ [fb]  & $K$ 
						& $\sigma$ [fb]  & $K$ & $\sigma$ [fb]  & $K$ 
						& $\sigma$ [fb]  & $K$ & $\sigma$ [fb]  & $K$ 
\tabularnewline\hline 
GF LO		& 65.4						& $\dots$ & 13.5				& $\dots$
		& 415						& $\dots$ & 115					& $\dots$	
		& 2.84$\times10^{3}$				& $\dots$ & 154					& $\dots$ \tabularnewline\hline	
GF NLL		& 65.9 $^{+  14 \%}_{ -26 \%}$			& 1.01	  & 13.7 $^{+  17 \%}_{ -27 \%}$	& 1.01
		& 414  $^{+   8 \%}_{ -23 \%}$			& 1.00	  & 115  $^{+  11 \%}_{ -24 \%}$	& 1.00	
		& 2.83 $^{+   2 \%}_{ -18 \%}\times10^{3}$	& 1.00	  & 154  $^{+   8 \%}_{ -21 \%}$	& 1.00 \tabularnewline\hline	
GF $\nkll{2}$	& 170  $^{   <1 \%}_{  -7 \%}$			& 2.61	  & 34.9 $^{   <1 \%}_{  -8 \%}$	& 2.59
		& 1.03 $^{   <1 \%}_{  -9 \%}\times10^{3}$	& 2.49	  & 281  $^{   <1 \%}_{  -7 \%}$  	& 2.45 
		& 6.83 $^{+   2 \%}_{ -13 \%}\times10^{3}$	& 2.40	  & 351  $^{   <1 \%}_{  -8 \%}$	& 2.28 \tabularnewline\hline	
GF $\nkll{3}$	& 202  $^{+   5 \%}_{ -11 \%}$			& 3.09	  & 41.3 $^{+   3 \%}_{  -9 \%}$	& 3.06
		& 1.21 $^{+   8 \%}_{ -13 \%}\times10^{3}$	& 2.92	  & 327  $^{+   6 \%}_{ -11 \%}$ 	& 2.85
		& 7.88 $^{+  13 \%}_{ -16 \%}\times10^{3}$	& 2.77	  & 401  $^{+   8 \%}_{ -11 \%}$	& 2.60 \tabularnewline
\end{tabular}
\caption{Heavy $N$ production cross sections via the GF mode at various accuracies, 
divided by active-heavy mixing $\vert V_{\ell N}\vert^2$, scale dependence $(\%)$, 
and $K$-factor, for representative $m_N$ and $\sqrt{s}$.
}
\label{tb:xSec}
\end{table*}
}

For the mass range under consideration, 
one observes unambiguously that the resummed GF rates at $\nkll{2}$ and $\nkll{3}$ are markedly larger than the LO rate, 
with $K\gtrsim\confirm{2-2.5}$ and notably independent of $m_N$.
This is unlike NLL, where $K\sim1$, since one essentially runs only $\alpha_s(\mu)$ and $C_\beta,~\tilde{s}_{\rm Higgs}\sim1$;
here, the uncertainty simply corresponds to varying $\alpha_s$.
We find $\sigma^{\nkll{3}} / \sigma^{\nkll{2}} \sim \confirm{1.1-1.2}$, indicating convergence of the perturbative series.

As previously stated, the residual uncertainty at $\nkll{3}$ stems from missing FO contributions.
Such terms, likely positive definite~\cite{Bonvini:2014qga}, 
consist of hard, initial state radiation (ISR) with $p_T^j \gtrsim \mu_f = Q$, 
which are not, by construction, included in the DGLAP-evolution of the PDFs.
The sizes of the $\nkll{2}$ and $\nkll{3}$ corrections are, in part, 
due to our scale choices and the desire to minimize the importance of missing QCD corrections.
Choosing alternative, less intuitive scales for the Born process can, of course, lead to smaller $K$-factors, but also to larger ones.
At both $\nkll{2}$ and $\nkll{3}$, the size of the uncertainty band is due to a residual $\mu_f$ dependence,
and requires matching to hard ISR from FO contributions to be reduced.
Moreover, these corrections are in line with those for Higgs and heavy (pseudo)scalar 
production~\cite{Dawson:1990zj,Spira:1995rr,Harlander:2002wh,Spira:1997dg,Harlander:2002vv,Anastasiou:2002wq,Anastasiou:2015ema,Ahmed:2016otz}.

In comparison with other heavy $N$ production modes, we find for $m_N\gtrsim\confirm{300}\GeV$ that the GF rate is now comparable to the CC and NC DY (not shown for clarity) rates. When basic fiducial cuts are applied on the charged lepton in the CC processes the combined GF+NC DY rate is slightly larger than the combined VBF+CC DY channel. For $m_N\lesssim600\GeV$, i.e., masses that are most relevant for LHC phenomenology due to mixing-suppression~\cite{Atre:2009rg,Alva:2014gxa}, the GF channel is factors larger than the VBF mechanism, indicating its potential importance at the LHC and its upgrades/successors.

We briefly note that the relative importance of the VBF mechanism found in Fig.~\ref{fig:heavyNXSec} is considerably less than what has been found in previous investigations, e.g., Ref.~\cite{Dev:2013wba} and follow-up works by the same authors. It was shown in Ref.~\cite{Alva:2014gxa} that the findings of Ref.~\cite{Dev:2013wba} were qualitatively and quantitatively incorrect: Their claimed ``$t$-channel enhancement'' are in reality due to several poorly/unregulated QCD and QED collinear divergences. Numerically, their cross sections were overestimated by $100\times$ in some instances. We refer readers to Refs.~\cite{Alva:2014gxa,Ruiz:2015zca,Degrande:2016aje} for correct, all-orders/resummed treatments of these contributions; to Ref.~\cite{Alva:2014gxa,Arganda:2015ija} for a quantitative assessment of $W\gamma$ scattering in heavy $N$ searches; and to Ref.~\cite{Degrande:2016aje} for non-expert-friendly infrared- and collinear-safe collider definitions for such processes.

In Fig.~\ref{fig:heavyNXSec}(R), we plot as a function of $\sqrt{s}$ for representative $m_N$
the summed GF + NC DY channels as well as the summed CC DY + VBF channels.
We add the channels incoherently, as GF is formally a noninterfering $\mathcal{O}(\alpha_s^2)$ correction to NC DY,
and similarly VBF is a noninterfering $\mathcal{O}(\alpha)$ correction to CC DY.
{Relative} uncertainties are added in quadrature.

We observe for \confirm{$m_N = 500-1000$} GeV
that the \textit{inclusive} production rate of $N\nu_\ell$ overtakes the \textit{inclusive} $N\ell^\pm$ production
at $\confirm{\sqrt{s} \gtrsim 15-30}$ TeV.
For $\sqrt{s}=33~(100)\TeV$, this difference is a factor of $\confirm{1-1.6~(2.5-2.7)}$ and is driven by the GF rate, 
for which the luminosity grows much faster than the $q\overline{q'}$ (DY) and $qq$ (VBF) luminosities with increasing collider energies.
While not shown, we find for $m_N = 500-1000\GeV$ 
that the GF rate individually exceeds the CC DY rate for $\sqrt{s}\confirm{\gtrsim20-25\TeV}$.
For increasing $\sqrt{s}$, we find that the resummation has a smaller impact on the total GF rate, with
\small
\begin{eqnarray}
\text{GF~}\nkll{3}	&:&  K=\confirm{2.6-3.6} ~\text{with}~ \delta\sigma/\sigma=\confirm{\pm 8-14  \%},\\
\text{GF~}\nkll{2}	&:&  K=\confirm{2.3-3.0} ~\text{with}~ \delta\sigma/\sigma=\confirm{\pm 6-11  \%},\\
\text{GF~}{\rm NLL}	&:&  K=\confirm{1.0-1.2} ~\text{with}~ \delta\sigma/\sigma=\confirm{\pm 19-29 \%}.\quad
\end{eqnarray}
\normalsize
This drop in $K$ is again due to an increasing importance of hard ISR,
and similarly leads to a sizable residual $\mu_f$ dependence.
In checks against heavy (pseudo)scalar production~\cite{Bonvini:2014joa,Bonvini:2014tea,Ahmed:2016otz},
we find similar results, and that the importance of FO corrections is $\mathcal{O}(+10\%)$.
Such corrections would likely push net $K$-factors for the $gg\to N\nu_\ell$ process to $K\sim3$, confirming the conjecture of~\cite{Hessler:2014ssa}.
We summarize our results in Table~\ref{tb:xSec}.

Due to the severe model-dependence of $V_{\ell N}$ as well as the associated phenomenology, 
an investigation into which is well beyond the scope of this study,
we defer further interpretation of our results to future studies.

\textbf{Usage:} For the use of these results in studies,
we advocate LO+parton shower event generation following~\cite{Degrande:2016aje}.
Total inclusive rates should then be normalized to those tabulated in Tables~\ref{tb:heavyNThresXSecAux1} and \ref{tb:heavyNThresXSecAux2}.
The flatness of the resummed $K$-factors means interpolation to unlisted $m_N$ is reliable.

\section{Summary and Conclusion}\label{sec:conclusions}
The existence of tiny neutrino masses and large mixing is unambiguous evidence for physics beyond the SM.
In light of Higgs boson data, the prospect of neutrino Dirac masses existing is increasingly likely.
Low-scale seesaw models with TeV-scale heavy neutrinos that couple appreciably to EW bosons 
are scenarios that can accommodate these seemingly contradictory observations, and still give rise to LHC phenomenology.

In this context, we have evaluated, for the first time, soft corrections to the GF production mode $gg \to Z^*/h^* \to N\nu_\ell$.
This was made possible by a new treatment  treatment of the $gg\to Z^*$ subprocess. 
For $m_N = 150-1000\GeV$ and $\sqrt{s} = 7 - 100$ TeV, we report:
\begin{eqnarray}
K^{\nkll{3}} = \sigma^{\nkll{3}}/\sigma^{\rm LO} &:&  \confirm{2.6-3.6},
  \\
K^{\nkll{2}} = \sigma^{\nkll{2}}/\sigma^{\rm LO} &:&  \confirm{2.3-3.0}.
\end{eqnarray}
We find that GF dominates over DY-like processes of Eq.~(\ref{eq:ccdy}) 
for \confirm{$m_N = 500-1000$} GeV at $\confirm{\sqrt{s} \gtrsim 20-25}$ TeV.
Corrections exhibit perturbative convergence and are consistent with Higgs and heavy (pseudo)scalar production.
Moreover, our results are independent of the precise nature/mixing of $N$,
and are expected to hold for other exotic leptons as well as other colorless, broken axial-vector currents one finds in other seesaw scenarios.

\noindent 
\acknowledgements{
\small{
\textit{
Acknowledgements: 
M.~Bonvini,
S.~Dawson,
C.~Fabrizio,
E.~Molinaro,
B.~Pecjak,
L.~Rottoli,
D.~Scott, 
C.-F.~Tamarit, and
C.~Weiland 
are thanked for discussions.
This work was funded in part by the UK STFC, and
the European Union's Horizon 2020 research and innovation programme under the Marie Sklodowska-Curie grant agreement 674896 (Elusives ITN).
RR acknowledges the CERN Theory group's generous hospitality during the completion of this work.
}}}

\appendix
\section{Appendix}
In Tables~\ref{tb:heavyNThresXSecAux1}-\ref{tb:heavyNThresXSecAux2}, we list
total hadronic cross sections for $gg\to Z^*/h^* \to N\nu_\ell$ at various accuracies, 
divided by active-heavy mixing $\vert V_{\ell N}\vert^2$, for representative collider energies $\sqrt{s} = 13,~14,~33,$ and 100 TeV.

\begin{table*}
 \renewcommand{\arraystretch}{1.8}
 \setlength\tabcolsep{6pt}
 \centering
 \begin{tabular}{ c || c | c | c | c }
 \hline\hline
 $m_N$	[GeV] &	$\sigma^{\rm LO}~/~ \vert V_{\ell N}\vert^2$ [fb]	& $\sigma^{\rm NLL}~/~ \vert V_{\ell N}\vert^2 ~$ [fb]	 
	      & $\sigma^{\rm N^2LL}~/~ \vert V_{\ell N}\vert^2 ~$ [fb] 	& $\sigma^{\rm N^3LL}~/~ \vert V_{\ell N}\vert^2 ~$ [fb]  \tabularnewline\hline
 \multicolumn{5}{c}{$\sqrt{s} = 13$ TeV}\tabularnewline\hline
  150  & 0.9097E+02  & 0.9131E+02 $^{+  11.8 \%}_{ -25.4 \%}$  & 0.2419E+03 $^{+   <0.1 \%}_{  -7.2 \%}$  & 0.2867E+03 $^{+   6.5 \%}_{ -12.3 \%}$   \tabularnewline\hline
  175  & 0.8811E+02  & 0.8868E+02 $^{+  12.4 \%}_{ -25.8 \%}$  & 0.2334E+03 $^{+   <0.1 \%}_{  -6.8 \%}$  & 0.2770E+03 $^{+   5.6 \%}_{ -12.0 \%}$   \tabularnewline\hline
  200  & 0.8510E+02  & 0.8560E+02 $^{+  12.9 \%}_{ -25.9 \%}$  & 0.2243E+03 $^{+   <0.1 \%}_{  -6.4 \%}$  & 0.2665E+03 $^{+   5.5 \%}_{ -11.6 \%}$   \tabularnewline\hline
  225  & 0.8020E+02  & 0.8075E+02 $^{+  13.3 \%}_{ -26.0 \%}$  & 0.2110E+03 $^{+   <0.1 \%}_{  -6.7 \%}$  & 0.2504E+03 $^{+   5.2 \%}_{ -11.3 \%}$   \tabularnewline\hline
  250  & 0.7330E+02  & 0.7379E+02 $^{+  13.7 \%}_{ -26.2 \%}$  & 0.1924E+03 $^{+   <0.1 \%}_{  -6.9 \%}$  & 0.2282E+03 $^{+   5.0 \%}_{ -11.1 \%}$   \tabularnewline\hline
  275  & 0.6451E+02  & 0.6496E+02 $^{+  14.1 \%}_{ -26.2 \%}$  & 0.1690E+03 $^{+   <0.1 \%}_{  -7.0 \%}$  & 0.2008E+03 $^{+   4.9 \%}_{ -10.8 \%}$   \tabularnewline\hline
  300  & 0.5467E+02  & 0.5507E+02 $^{+  14.6 \%}_{ -26.4 \%}$  & 0.1430E+03 $^{+   <0.1 \%}_{  -7.2 \%}$  & 0.1697E+03 $^{+   4.6 \%}_{ -10.4 \%}$   \tabularnewline\hline
  325  & 0.4500E+02  & 0.4515E+02 $^{+  15.0 \%}_{ -26.3 \%}$  & 0.1171E+03 $^{+   <0.1 \%}_{  -7.2 \%}$  & 0.1388E+03 $^{+   4.7 \%}_{ -10.2 \%}$   \tabularnewline\hline
  350  & 0.3629E+02  & 0.3640E+02 $^{+  15.6 \%}_{ -26.5 \%}$  & 0.9423E+02 $^{+   <0.1 \%}_{  -7.5 \%}$  & 0.1117E+03 $^{+   4.4 \%}_{  -9.9 \%}$   \tabularnewline\hline
  375  & 0.2922E+02  & 0.2947E+02 $^{+  15.6 \%}_{ -26.7 \%}$  & 0.7613E+02 $^{+   <0.1 \%}_{  -7.7 \%}$  & 0.9010E+02 $^{+   4.2 \%}_{  -9.7 \%}$   \tabularnewline\hline
  400  & 0.2370E+02  & 0.2396E+02 $^{+  16.1 \%}_{ -26.7 \%}$  & 0.6185E+02 $^{+   <0.1 \%}_{  -7.8 \%}$  & 0.7328E+02 $^{+   3.9 \%}_{  -9.6 \%}$   \tabularnewline\hline
  450  & 0.1593E+02  & 0.1612E+02 $^{+  16.8 \%}_{ -27.0 \%}$  & 0.4147E+02 $^{+   <0.1 \%}_{  -8.2 \%}$  & 0.4910E+02 $^{+   3.4 \%}_{  -9.3 \%}$   \tabularnewline\hline
  500  & 0.1089E+02  & 0.1106E+02 $^{+  17.2 \%}_{ -27.1 \%}$  & 0.2838E+02 $^{+   <0.1 \%}_{  -8.4 \%}$  & 0.3355E+02 $^{+   3.4 \%}_{  -8.9 \%}$   \tabularnewline\hline
  550  & 0.7617E+01  & 0.7747E+01 $^{+  17.5 \%}_{ -27.3 \%}$  & 0.1984E+02 $^{+   <0.1 \%}_{  -8.6 \%}$  & 0.2351E+02 $^{+   3.2 \%}_{  -8.8 \%}$   \tabularnewline\hline
  600  & 0.5413E+01  & 0.5518E+01 $^{+  18.0 \%}_{ -27.3 \%}$  & 0.1410E+02 $^{+   <0.1 \%}_{  -8.7 \%}$  & 0.1670E+02 $^{+   3.0 \%}_{  -8.5 \%}$   \tabularnewline\hline
 \multicolumn{5}{c}{$\sqrt{s} = 14$ TeV}\tabularnewline\hline
  150  & 0.1065E+03  & 0.1069E+03 $^{+  11.2 \%}_{ -25.0 \%}$  & 0.2824E+03 $^{+   <0.1 \%}_{  -7.5 \%}$  & 0.3348E+03 $^{+   6.7 \%}_{ -12.7 \%}$   \tabularnewline\hline
  175  & 0.1039E+03  & 0.1043E+03 $^{+  11.8 \%}_{ -25.3 \%}$  & 0.2733E+03 $^{+   <0.1 \%}_{  -7.2 \%}$  & 0.3239E+03 $^{+   6.1 \%}_{ -12.2 \%}$   \tabularnewline\hline
  200  & 0.1006E+03  & 0.1012E+03 $^{+  12.4 \%}_{ -25.5 \%}$  & 0.2641E+03 $^{+   <0.1 \%}_{  -6.7 \%}$  & 0.3132E+03 $^{+   5.7 \%}_{ -11.9 \%}$   \tabularnewline\hline
  225  & 0.9522E+02  & 0.9567E+02 $^{+  13.0 \%}_{ -25.7 \%}$  & 0.2490E+03 $^{+   <0.1 \%}_{  -6.3 \%}$  & 0.2952E+03 $^{+   5.4 \%}_{ -11.5 \%}$   \tabularnewline\hline
  250  & 0.8711E+02  & 0.8776E+02 $^{+  13.2 \%}_{ -25.9 \%}$  & 0.2279E+03 $^{+   <0.1 \%}_{  -6.7 \%}$  & 0.2702E+03 $^{+   5.1 \%}_{ -11.3 \%}$   \tabularnewline\hline
  275  & 0.7693E+02  & 0.7746E+02 $^{+  13.6 \%}_{ -25.9 \%}$  & 0.2008E+03 $^{+   <0.1 \%}_{  -6.7 \%}$  & 0.2380E+03 $^{+   5.1 \%}_{ -11.0 \%}$   \tabularnewline\hline
  300  & 0.6538E+02  & 0.6585E+02 $^{+  13.9 \%}_{ -26.0 \%}$  & 0.1704E+03 $^{+   <0.1 \%}_{  -6.8 \%}$  & 0.2018E+03 $^{+   5.0 \%}_{ -10.7 \%}$   \tabularnewline\hline
  325  & 0.5386E+02  & 0.5425E+02 $^{+  14.4 \%}_{ -26.1 \%}$  & 0.1401E+03 $^{+   <0.1 \%}_{  -7.0 \%}$  & 0.1659E+03 $^{+   4.7 \%}_{ -10.5 \%}$   \tabularnewline\hline
  350  & 0.4377E+02  & 0.4395E+02 $^{+  14.9 \%}_{ -26.2 \%}$  & 0.1133E+03 $^{+   <0.1 \%}_{  -7.1 \%}$  & 0.1341E+03 $^{+   4.6 \%}_{ -10.2 \%}$   \tabularnewline\hline
  375  & 0.3557E+02  & 0.3565E+02 $^{+  15.5 \%}_{ -26.2 \%}$  & 0.9174E+02 $^{+   <0.1 \%}_{  -7.2 \%}$  & 0.1085E+03 $^{+   4.5 \%}_{  -9.8 \%}$   \tabularnewline\hline
  400  & 0.2900E+02  & 0.2913E+02 $^{+  15.6 \%}_{ -26.5 \%}$  & 0.7480E+02 $^{+   <0.1 \%}_{  -7.5 \%}$  & 0.8841E+02 $^{+   4.2 \%}_{  -9.7 \%}$   \tabularnewline\hline
  450  & 0.1951E+02  & 0.1975E+02 $^{+  16.2 \%}_{ -26.7 \%}$  & 0.5061E+02 $^{+   <0.1 \%}_{  -7.9 \%}$  & 0.5985E+02 $^{+   3.8 \%}_{  -9.5 \%}$   \tabularnewline\hline
  500  & 0.1351E+02  & 0.1367E+02 $^{+  16.8 \%}_{ -26.9 \%}$  & 0.3493E+02 $^{+   <0.1 \%}_{  -8.1 \%}$  & 0.4128E+02 $^{+   3.4 \%}_{  -9.2 \%}$   \tabularnewline\hline
  550  & 0.9494E+01  & 0.9636E+01 $^{+  17.2 \%}_{ -27.0 \%}$  & 0.2456E+02 $^{+   <0.1 \%}_{  -8.3 \%}$  & 0.2899E+02 $^{+   3.5 \%}_{  -8.8 \%}$   \tabularnewline\hline
  600  & 0.6807E+01  & 0.6922E+01 $^{+  17.4 \%}_{ -27.1 \%}$  & 0.1762E+02 $^{+   <0.1 \%}_{  -8.5 \%}$  & 0.2083E+02 $^{+   3.2 \%}_{  -8.7 \%}$   \tabularnewline\hline
 \hline
\end{tabular}
\caption{Total hadronic cross sections for $gg\to Z^*/h^* \to N\nu_\ell$ at various accuracies, 
divided by active-heavy mixing $\vert V_{\ell N}\vert^2$, for representative collider energies $\sqrt{s}$.}
\label{tb:heavyNThresXSecAux1}
\end{table*}

\begin{table*}
 \renewcommand{\arraystretch}{1.8}
 \setlength\tabcolsep{6pt}
 \centering
 \begin{tabular}{ c || c | c | c | c }
 \hline\hline
 $m_N$	[GeV] &	$\sigma^{\rm LO}~/~ \vert V_{\ell N}\vert^2$ [fb]	& $\sigma^{\rm NLL}~/~ \vert V_{\ell N}\vert^2 ~$ [fb]	 
	      & $\sigma^{\rm N^2LL}~/~ \vert V_{\ell N}\vert^2 ~$ [fb] 	& $\sigma^{\rm N^3LL}~/~ \vert V_{\ell N}\vert^2 ~$ [fb]  \tabularnewline\hline
 \multicolumn{4}{c}{$\sqrt{s} = 33$ TeV}\tabularnewline\hline
  150  & 0.5547E+03  & 0.5510E+03 $^{+   5.8 \%}_{ -21.3 \%}$  & 0.1408E+04 $^{+   <0.1 \%}_{ -10.8 \%}$  & 0.1658E+04 $^{+  10.3 \%}_{ -15.1 \%}$   \tabularnewline\hline
  200  & 0.5672E+03  & 0.5663E+03 $^{+   6.3 \%}_{ -21.9 \%}$  & 0.1428E+04 $^{+   <0.1 \%}_{ -10.2 \%}$  & 0.1674E+04 $^{+   9.7 \%}_{ -14.1 \%}$   \tabularnewline\hline
  300  & 0.4146E+03  & 0.4144E+03 $^{+   8.3 \%}_{ -22.5 \%}$  & 0.1033E+04 $^{+   <0.1 \%}_{  -8.8 \%}$  & 0.1209E+04 $^{+   8.3 \%}_{ -13.1 \%}$   \tabularnewline\hline
  400  & 0.2132E+03  & 0.2127E+03 $^{+  10.0 \%}_{ -23.1 \%}$  & 0.5250E+03 $^{+   <0.1 \%}_{  -7.4 \%}$  & 0.6134E+03 $^{+   7.1 \%}_{ -11.9 \%}$   \tabularnewline\hline
  500  & 0.1147E+03  & 0.1146E+03 $^{+  11.1 \%}_{ -23.6 \%}$  & 0.2806E+03 $^{+   <0.1 \%}_{  -6.6 \%}$  & 0.3273E+03 $^{+   6.3 \%}_{ -11.1 \%}$   \tabularnewline\hline
  600  & 0.6606E+02  & 0.6615E+02 $^{+  12.0 \%}_{ -23.9 \%}$  & 0.1608E+03 $^{+   <0.1 \%}_{  -5.9 \%}$  & 0.1873E+03 $^{+   5.6 \%}_{ -10.5 \%}$   \tabularnewline\hline
  700  & 0.4011E+02  & 0.4025E+02 $^{+  13.0 \%}_{ -24.2 \%}$  & 0.9747E+02 $^{+   <0.1 \%}_{  -5.5 \%}$  & 0.1135E+03 $^{+   5.2 \%}_{  -9.9 \%}$   \tabularnewline\hline
  800  & 0.2550E+02  & 0.2564E+02 $^{+  13.6 \%}_{ -24.4 \%}$  & 0.6183E+02 $^{+   <0.1 \%}_{  -5.9 \%}$  & 0.7191E+02 $^{+   4.8 \%}_{  -9.5 \%}$   \tabularnewline\hline
  900  & 0.1681E+02  & 0.1690E+02 $^{+  14.2 \%}_{ -24.7 \%}$  & 0.4066E+02 $^{+   <0.1 \%}_{  -6.3 \%}$  & 0.4733E+02 $^{+   4.4 \%}_{  -9.3 \%}$   \tabularnewline\hline
 1000  & 0.1141E+02  & 0.1147E+02 $^{+  14.8 \%}_{ -24.7 \%}$  & 0.2753E+02 $^{+   <0.1 \%}_{  -6.4 \%}$  & 0.3204E+02 $^{+   4.2 \%}_{  -9.0 \%}$   \tabularnewline\hline
 1100  & 0.7916E+01  & 0.7981E+01 $^{+  15.2 \%}_{ -24.9 \%}$  & 0.1912E+02 $^{+   <0.1 \%}_{  -6.6 \%}$  & 0.2228E+02 $^{+   4.0 \%}_{  -8.7 \%}$   \tabularnewline\hline
 1200  & 0.5625E+01  & 0.5685E+01 $^{+  15.4 \%}_{ -25.0 \%}$  & 0.1360E+02 $^{+   <0.1 \%}_{  -6.8 \%}$  & 0.1582E+02 $^{+   3.8 \%}_{  -8.6 \%}$   \tabularnewline\hline
 1300  & 0.4066E+01  & 0.4115E+01 $^{+  15.8 \%}_{ -25.1 \%}$  & 0.9831E+01 $^{+   <0.1 \%}_{  -7.0 \%}$  & 0.1144E+02 $^{+   3.7 \%}_{  -8.4 \%}$   \tabularnewline\hline
 1400  & 0.2985E+01  & 0.3031E+01 $^{+  16.0 \%}_{ -25.2 \%}$  & 0.7234E+01 $^{+   <0.1 \%}_{  -7.1 \%}$  & 0.8425E+01 $^{+   3.4 \%}_{  -8.4 \%}$   \tabularnewline\hline
 1500  & 0.2225E+01  & 0.2261E+01 $^{+  16.3 \%}_{ -25.2 \%}$  & 0.5389E+01 $^{+   <0.1 \%}_{  -7.1 \%}$  & 0.6281E+01 $^{+   3.5 \%}_{  -8.2 \%}$   \tabularnewline\hline
 \multicolumn{4}{c}{$\sqrt{s} = 100$ TeV}\tabularnewline\hline
  150  & 0.3230E+04  & 0.3223E+04 $^{+  <0.1 \%}_{ -16.8 \%}$  & 0.7967E+04 $^{+   3.9 \%}_{ -15.5 \%}$  & 0.9244E+04 $^{+  15.2 \%}_{ -18.8 \%}$   \tabularnewline\hline
  200  & 0.3516E+04  & 0.3507E+04 $^{+   0.1 \%}_{ -17.2 \%}$  & 0.8569E+04 $^{+   3.2 \%}_{ -14.3 \%}$  & 0.9922E+04 $^{+  14.1 \%}_{ -17.6 \%}$   \tabularnewline\hline
  300  & 0.2839E+04  & 0.2832E+04 $^{+   1.8 \%}_{ -17.9 \%}$  & 0.6825E+04 $^{+   2.2 \%}_{ -12.9 \%}$  & 0.7876E+04 $^{+  13.0 \%}_{ -16.2 \%}$   \tabularnewline\hline
  400  & 0.1648E+04  & 0.1645E+04 $^{+   3.3 \%}_{ -18.7 \%}$  & 0.3909E+04 $^{+   1.0 \%}_{ -11.6 \%}$  & 0.4493E+04 $^{+  11.6 \%}_{ -15.0 \%}$   \tabularnewline\hline
  500  & 0.9911E+03  & 0.9913E+03 $^{+   4.7 \%}_{ -19.2 \%}$  & 0.2330E+04 $^{+   0.2 \%}_{ -10.5 \%}$  & 0.2674E+04 $^{+  10.5 \%}_{ -14.0 \%}$   \tabularnewline\hline
  600  & 0.6330E+03  & 0.6322E+03 $^{+   5.7 \%}_{ -19.7 \%}$  & 0.1473E+04 $^{+   <0.1 \%}_{  -9.6 \%}$  & 0.1688E+04 $^{+   9.8 \%}_{ -13.1 \%}$   \tabularnewline\hline
  700  & 0.4242E+03  & 0.4220E+03 $^{+   6.6 \%}_{ -20.0 \%}$  & 0.9775E+03 $^{+   <0.1 \%}_{  -8.9 \%}$  & 0.1119E+04 $^{+   9.0 \%}_{ -12.6 \%}$   \tabularnewline\hline
  800  & 0.2937E+03  & 0.2938E+03 $^{+   7.3 \%}_{ -20.4 \%}$  & 0.6768E+03 $^{+   <0.1 \%}_{  -8.4 \%}$  & 0.7728E+03 $^{+   8.3 \%}_{ -11.9 \%}$   \tabularnewline\hline
  900  & 0.2101E+03  & 0.2101E+03 $^{+   7.8 \%}_{ -20.5 \%}$  & 0.4817E+03 $^{+   <0.1 \%}_{  -8.0 \%}$  & 0.5498E+03 $^{+   8.1 \%}_{ -11.6 \%}$   \tabularnewline\hline
 1000  & 0.1541E+03  & 0.1538E+03 $^{+   8.4 \%}_{ -20.8 \%}$  & 0.3513E+03 $^{+   <0.1 \%}_{  -7.5 \%}$  & 0.4010E+03 $^{+   7.8 \%}_{ -11.2 \%}$   \tabularnewline\hline
 1100  & 0.1152E+03  & 0.1153E+03 $^{+   9.0 \%}_{ -21.0 \%}$  & 0.2624E+03 $^{+   <0.1 \%}_{  -7.1 \%}$  & 0.2996E+03 $^{+   7.5 \%}_{ -10.9 \%}$   \tabularnewline\hline
 1200  & 0.8800E+02  & 0.8830E+02 $^{+   9.1 \%}_{ -21.3 \%}$  & 0.2006E+03 $^{+   <0.1 \%}_{  -7.1 \%}$  & 0.2291E+03 $^{+   6.9 \%}_{ -11.0 \%}$   \tabularnewline\hline
 1300  & 0.6822E+02  & 0.6839E+02 $^{+   9.8 \%}_{ -21.4 \%}$  & 0.1550E+03 $^{+   <0.1 \%}_{  -6.5 \%}$  & 0.1771E+03 $^{+   6.6 \%}_{ -10.4 \%}$   \tabularnewline\hline
 1400  & 0.5369E+02  & 0.5371E+02 $^{+  10.1 \%}_{ -21.5 \%}$  & 0.1214E+03 $^{+   <0.1 \%}_{  -6.2 \%}$  & 0.1384E+03 $^{+   6.6 \%}_{ -10.1 \%}$   \tabularnewline\hline
 1500  & 0.4272E+02  & 0.4278E+02 $^{+  10.4 \%}_{ -21.7 \%}$  & 0.9653E+02 $^{+   <0.1 \%}_{  -6.0 \%}$  & 0.1102E+03 $^{+   6.2 \%}_{ -10.1 \%}$   \tabularnewline\hline
 \hline
\end{tabular}
\caption{Total hadronic cross sections for $gg\to Z^*/h^* \to N\nu_\ell$ at various accuracies,
divided by active-heavy mixing $\vert V_{\ell N}\vert^2$, for representative collider energies $\sqrt{s}$.}
\label{tb:heavyNThresXSecAux2}
\end{table*}

\newpage


\end{document}